# Absolute Properties of the Eclipsing Binary Star AP Andromedae

Claud H. Sandberg Lacy[1] Guillermo Torres[2], Francis C. Fekel[3], Matthew W. Muterspaugh[3]

[1]Physics Department, University of Arkansas, Fayetteville, AR 72701, USA; clacy@uark.edu

[2]Harvard-Smithsonian Center for Astrophysics, 60 Garden Street, Cambridge, MA 02138;

gtorres@cfa.harvard.edu

[3]Center of Excellence in Information Systems, Tennessee State University, Nashville, TN 37209, USA;

fekel@evans.tsuniv.edu, matthew1@coe.tsuniv.edu

AP And is a well-detached F5 eclipsing binary star for which only a very limited amount of information was available before this publication. We have obtained very extensive measurements of the light curve (19097 differential V magnitude observations) and a radial velocity curve (83 spectroscopic observations) which allow us to fit orbits and determine the absolute properties of the components very accurately: masses of $1.277 \pm 0.004$ and $1.251 \pm 0.004$ solar masses, radii of $1.233 \pm 0.006$ and $1.1953 \pm 0.005$ solar radii, and temperatures of $6565 \pm 150$ K and $6495 \pm 150$ K. The distance to the system is about $400 \pm 30$ pc. Comparison with the theoretical properties of the stellar evolutionary models of the Yonsei-Yale series of Yi et al. shows good agreement between the observations and the theory at an age of about 500 Myr and a slightly sub-solar metallicity.

1. Introduction

AP And (classified EA/DM, mag. 11.30-11.90, TYC 3639-0915-1, spectral type F5) was found to be an eclipsing binary star by B.S. Whitney (1957). He determined the first accurate eclipse ephemeris based on his photographic data. Other than numerous times of minimum light, little else has been published on this star until now. We began photometric observations by using robotic telescopes (the URSA and NFO WebScopes) in 2003 July, eventually accumulating 19097 measurable images, from which accurate differential magnitudes could be determined. Spectroscopic observations began in 2004 with CfA telescopes, which have now accumulated 42 spectra. These are supplemented by 41 spectra gathered by a robotic telescope at Fairborn Observatory. The eclipse ephemeris is discussed in Sec. 2, the spectroscopic observations and reductions in Sec. 3, the spectroscopic orbital solution in Sec. 4, and the photometric observations and orbit in Sec. 5. The combination of all the existing data have allowed us to determine very accurate absolute properties of the stars in this system (Sec. 6), and to compare our observations with the standard theory of stellar evolution, which we find agrees well with our observed results at an age of about 500 Myr.

2. Eclipse ephemeris

We have gathered the existing accurate photoelectric and CCD dates of minima to find the current eclipse ephemeris (Table 1 below, and the data are also available in a machine-readable version with the electronic edition of this journal), which we determine to be:

$$\text{HJD Min I} = 2{,}454{,}717.65759(2) + 1.587291156(33) \, E \tag{1}$$

where the uncertainty in the last digits of the period and epoch is shown in parentheses. All uncertainties quoted in this paper are standard errors unless otherwise mentioned. We have rescaled them here in order to produce a reduced chi-square value of unity in the Levenberg-Marquardt minimization procedure we used, separately for the type 1 and type 2 eclipses since those measurements have somewhat different precisions. This rebalances the corresponding weights and yields more accurate estimates for the orbital period and epoch and their corresponding uncertainties. The phase of secondary eclipse is found to be 0.50000 ± 0.00003, so no eccentricity is apparent in the orbit. The minima that were used in the fit and their residuals are shown in Table 1. Less accurate photographic and visual dates in the literature were checked against this fitted ephemeris, but no trends in the older residuals were seen.

Table 1. Observed dates of minimum light for AP And

| HJD-2400000 | Type[a] | Uncertainty (d) | Residual O-C (d) | Ref |
|---|---|---|---|---|
| 52530.37050 | 1 | 0.00020 | 0.00012 | 1 |
| 53006.55780 | 1 | 0.00010 | 0.00008 | 2 |
| 53201.79472 | 1 | 0.00015 | 0.00018 | 3 |
| 53205.76350 | 2 | 0.00020 | 0.00074 | 3 |
| 53220.84200 | 1 | 0.00010 | -0.00003 | 3 |
| 53302.58670 | 2 | 0.00040 | -0.00082 | 3 |
| 53348.61890 | 2 | 0.00010 | -0.00007 | 4 |
| 53349.41390 | 1 | 0.00070 | 0.00129 | 5 |
| 53411.31680 | 1 | 0.00090 | -0.00017 | 5 |
| 53574.80780 | 1 | 0.00020 | -0.00016 | 4 |
| 53618.45860 | 2 | 0.00040 | 0.00014 | 6 |
| 53631.95040 | 1 | 0.00010 | -0.00004 | 4 |
| 53660.52130 | 1 | 0.00020 | -0.00038 | 6 |
| 53670.83898 | 2 | 0.00006 | -0.00009 | 4 |
| 53671.63270 | 1 | 0.00010 | -0.00002 | 7 |
| 53671.63280 | 1 | 0.00010 | 0.00008 | 4 |
| 53698.61650 | 1 | 0.00010 | -0.00017 | 4 |
| 53706.55320 | 1 | 0.00010 | 0.00008 | 4 |
| 53733.53720 | 1 | 0.00010 | 0.00013 | 8 |
| 53736.71190 | 1 | 0.00040 | 0.00024 | 8 |
| 53916.86940 | 2 | 0.00030 | 0.00020 | 8 |
| 53998.61470 | 1 | 0.00020 | 0.00000 | 8 |
| 54009.72560 | 1 | 0.00020 | -0.00013 | 8 |
| 54017.66190 | 1 | 0.00020 | -0.00029 | 8 |
| 54017.66360 | 1 | 0.00050 | 0.00141 | 9 |
| 54021.63020 | 2 | 0.00020 | -0.00022 | 8 |
| 54021.63030 | 2 | 0.00020 | -0.00012 | 8 |
| 54026.39030 | 2 | 0.00060 | -0.00199 | 9 |
| 54028.77330 | 1 | 0.00010 | 0.00007 | 8 |
| 54029.56700 | 2 | 0.00010 | 0.00013 | 8 |
| 54032.74140 | 2 | 0.00010 | -0.00006 | 8 |
| 54048.61430 | 2 | 0.00020 | -0.00007 | 8 |
| 54051.78920 | 2 | 0.00020 | 0.00025 | 8 |
| 54052.58240 | 1 | 0.00020 | -0.00020 | 8 |
| 54059.72530 | 2 | 0.00020 | -0.00011 | 8 |
| 54063.69390 | 1 | 0.00010 | 0.00027 | 8 |
| 54067.66200 | 2 | 0.00030 | 0.00014 | 8 |
| 54071.62990 | 1 | 0.00020 | -0.00019 | 8 |
| 54071.63040 | 1 | 0.00030 | 0.00031 | 8 |
| 54075.59820 | 2 | 0.00030 | -0.00012 | 8 |
| 54082.74100 | 1 | 0.00010 | -0.00013 | 8 |
| 54086.70910 | 2 | 0.00010 | -0.00026 | 8 |
| 54094.64580 | 2 | 0.00010 | -0.00001 | 8 |
| 54110.51860 | 2 | 0.00020 | -0.00012 | 8 |
| 54285.91430 | 1 | 0.00020 | -0.00010 | 10 |
| 54289.88260 | 2 | 0.00030 | -0.00002 | 10 |
| 54317.66010 | 1 | 0.00040 | -0.00012 | 10 |
| 54327.97780 | 2 | 0.00020 | 0.00019 | 10 |
| 54328.77080 | 1 | 0.00020 | -0.00046 | 10 |
| 54328.77130 | 1 | 0.00020 | 0.00004 | 10 |
| 54331.94560 | 1 | 0.00020 | −0.00024 | 10 |
| 54339.88230 | 1 | 0.00020 | 0.00001 | 10 |
| 54339.88300 | 1 | 0.00040 | 0.00071 | 10 |
| 54343.85100 | 2 | 0.00020 | 0.00048 | 10 |
| 54347.81940 | 1 | 0.00050 | 0.00065 | 10 |
| 54360.51740 | 1 | 0.00080 | 0.00032 | 11 |
| 54367.66010 | 2 | 0.00010 | 0.00021 | 10 |
| 54371.62810 | 1 | 0.00020 | −0.00002 | 10 |
| 54386.70740 | 2 | 0.00020 | 0.00002 | 10 |
| 54389.88240 | 2 | 0.00020 | 0.00043 | 10 |
| 54393.85030 | 1 | 0.00020 | 0.00011 | 10 |
| 54394.64400 | 2 | 0.00020 | 0.00016 | 10 |
| 54398.61220 | 1 | 0.00020 | 0.00013 | 10 |
| 54401.78650 | 1 | 0.00020 | −0.00015 | 10 |
| 54401.78670 | 1 | 0.00020 | 0.00005 | 10 |
| 54405.75500 | 2 | 0.00010 | 0.00012 | 10 |
| 54409.72290 | 1 | 0.00020 | −0.00021 | 10 |
| 54413.69110 | 2 | 0.00020 | −0.00023 | 10 |
| 54413.69140 | 2 | 0.00040 | 0.00007 | 10 |
| 54421.62780 | 2 | 0.00020 | 0.00001 | 10 |
| 54459.72260 | 2 | 0.00020 | −0.00018 | 10 |
| 54463.69090 | 1 | 0.00010 | −0.00011 | 10 |
| 54475.59540 | 2 | 0.00020 | −0.00029 | 10 |
| 54498.61130 | 1 | 0.00020 | −0.00011 | 10 |
| 55097.81420 | 2 | 0.00020 | 0.00038 | 10 |
| 55121.62290 | 2 | 0.00010 | −0.00029 | 12 |
| 55139.87690 | 1 | 0.00050 | −0.00014 | 12 |
| 55144.63910 | 1 | 0.00010 | 0.00019 | 12 |
| 55144.64060 | 1 | 0.00030 | 0.00169 | 13 |
| 55152.57550 | 1 | 0.00010 | 0.00013 | 12 |
| 55159.71790 | 2 | 0.00020 | −0.00028 | 12 |
| 55358.92300 | 1 | 0.00020 | −0.00022 | 12 |
| 55412.89110 | 1 | 0.00010 | −0.00002 | 12 |
| 55432.73160 | 2 | 0.00020 | −0.00066 | 12 |
| 55451.77960 | 2 | 0.00010 | −0.00015 | 12 |
| 55466.85870 | 1 | 0.00020 | −0.00032 | 12 |
| 55467.65310 | 2 | 0.00020 | 0.00044 | 12 |
| 55478.76420 | 2 | 0.00030 | 0.00050 | 12 |
| 55486.70030 | 2 | 0.00010 | 0.00014 | 12 |
| 55491.46250 | 2 | 0.00060 | 0.00047 | 14 |
| 55494.63600 | 2 | 0.00020 | −0.00061 | 12 |
| 55497.01760 | 1 | 0.00100 | 0.00005 | 15 |
| 55497.81120 | 2 | 0.00020 | 0.00001 | 12 |
| 55509.71560 | 1 | 0.00020 | −0.00028 | 12 |
| 55513.68450 | 2 | 0.00020 | 0.00040 | 16 |
| 55528.76330 | 1 | 0.00020 | −0.00007 | 12 |
| 55555.74700 | 1 | 0.00020 | −0.00032 | 12 |
| 55563.68370 | 1 | 0.00010 | −0.00008 | 12 |
| 55575.58840 | 2 | 0.00020 | −0.00006 | 12 |
| 55575.58850 | 2 | 0.00020 | 0.00004 | 12 |
| 55743.84090 | 2 | 0.00020 | −0.00042 | 17 |
| 55758.92060 | 1 | 0.00030 | 0.00001 | 17 |
| 55770.82540 | 2 | 0.00020 | 0.00013 | 17 |
| 55801.77720 | 1 | 0.00020 | −0.00025 | 17 |
| 55809.71380 | 1 | 0.00010 | −0.00011 | 17 |
| 55816.85620 | 2 | 0.00030 | −0.00052 | 17 |
| 55824.79310 | 2 | 0.00010 | −0.00007 | 17 |
| 55835.90460 | 2 | 0.00030 | 0.00039 | 17 |
| 55837.49160 | 2 | 0.00010 | 0.00010 | 18 |
| 55838.28510 | 1 | 0.00020 | −0.00005 | 18 |
| 55841.45970 | 1 | 0.00010 | −0.00003 | 18 |
| 55848.60230 | 2 | 0.00020 | −0.00024 | 17 |
| 55848.60290 | 2 | 0.00020 | 0.00036 | 17 |
| 55851.77690 | 2 | 0.00020 | −0.00022 | 17 |
| 55851.77720 | 2 | 0.00010 | 0.00008 | 17 |
| 55853.36460 | 2 | 0.00010 | 0.00019 | 18 |
| 55856.53920 | 2 | 0.00030 | 0.00021 | 17 |
| 55866.85660 | 1 | 0.00010 | 0.00021 | 17 |
| 55875.58670 | 2 | 0.00030 | 0.00021 | 17 |
| 55890.66580 | 1 | 0.00020 | 0.00005 | 17 |
| 55894.63390 | 2 | 0.00010 | −0.00008 | 17 |
| 55925.58640 | 1 | 0.00020 | 0.00024 | 17 |
| 56085.90220 | 1 | 0.00030 | −0.00037 | 19 |
| 56120.82270 | 1 | 0.00020 | −0.00027 | 19 |
| 56186.69570 | 2 | 0.00020 | 0.00015 | 19 |
| 56190.66350 | 1 | 0.00030 | −0.00028 | 19 |
| 56225.58450 | 1 | 0.00030 | 0.00031 | 19 |
| 56298.59930 | 1 | 0.00020 | −0.00028 | 19 |
| 56566.85180 | 1 | 0.00010 | 0.00001 | 20 |
| 56567.64540 | 2 | 0.00020 | −0.00003 | 20 |
| 56586.69270 | 2 | 0.00030 | −0.00023 | 20 |
| 56593.83610 | 1 | 0.00030 | 0.00036 | 20 |
| 56597.80460 | 2 | 0.00020 | 0.00064 | 20 |
| 56598.59760 | 1 | 0.00020 | −0.00001 | 20 |
| 56628.75600 | 1 | 0.00050 | −0.00014 | 20 |

**References:**

(1) Agerer & Hübscher 2003; (2) Dvorak 2005; (3) Lacy 2004; (4) Lacy 2006; (5) Hübscher et al. 2005; (6) Hübscher et al. 2006; (7) Nelson 2006; (8) Lacy 2007; (9) Hübscher & Walter 2007; (10) Lacy 2009; (11) Hübscher et al. 2008; (12) Lacy 2011; (13) Diethlem 2010; (14) Hübscher 2011; (15) Nagai 2011; (16) Diethlem 2011; (17) Lacy 2012; (18) Liakos & Niarchos 2011; (19) Lacy 2013; (20) this paper.

[a]Eclipses of type 1 are the deeper eclipses when the hotter, more massive star (star A) is being eclipsed by the cooler, less massive star (star B).

3. Spectroscopic observations and reductions

AP And was monitored spectroscopically with three different instruments over more than nine years. Observations began at the Harvard-Smithsonian Center for Astrophysics (CfA) with the 1.5m telescope at the F. L. Whipple Observatory (Mount Hopkins, AZ). A single echelle order 45 Å wide centered at about 5187 Å (Mg I b triplet) was recorded with an intensified photon-counting Reticon detector (Digital Speedometer, DS, Latham 1992), with a resolving power of approximately R = 35,000. Signal-to-noise ratios range from 21 to 33 per resolution element of 8.5 km/s. A total of 16 spectra were collected between 2004 January and 2008 October.

Additional observations were gathered on the same telescope using the bench-mounted Tillinghast Reflector Echelle Spectrograph (TRES, Furesz 2008), from 2009 October to 2013 January. The resolving power of this instrument is R = 44,000, and the wavelength coverage is 3900-9100 Å in 51 orders. The signal-to-noise ratios of the 26 spectra we obtained range from 19 to 74 per resolution element of 6.8 km/s.

A further 41 echelle spectrograms of AP And were collected from 2011 October to 2013 October using the Tennessee State University 2m telescope and a fiber fed echelle spectrograph (Eaton & Williamson 2007). The detector was a Fairchild 486 CCD with a 4096 x 4096 array of 15 micron pixels. The resulting echelle spectrograms have 48 orders and have a wavelength coverage of 3800-8260 Å. Because of the star's faintness, we used a fiber that produced a resolution of 0.4 Å, or a resolving power of 15,000 at 6000 Å. The typical signal-to-noise ratio of these spectra is 45 at 6000 Å.

Radial velocities for the CfA spectra (DS and TRES) were measured using the two-dimensional cross-correlation algorithm TODCOR (Zucker & Mazeh 1994). In the case of TRES we used only order 23 (centered on the Mg I b triplet), which contains most of the velocity information. Templates were selected from a large library of synthetic spectra based on model atmospheres by R. L. Kurucz (Nordstrom et al. 1994; Latham et al. 2002), available for a wide range of temperatures ($T_{eff}$), surface gravities (log g), metallicities ([Fe/H]), and rotational velocities ($v_{rot}$ sin i when seen in projection). The optimum template parameters were chosen by cross-correlating all observations against the entire library of synthetic spectra, seeking the highest peak in the cross-correlation function averaged over all exposures (see Torres et al. 2002). This was done separately for the DS and TRES spectra. In each case we held log g fixed at the value 4.5, close to our final estimates in Sect. 5, and we assumed solar metallicity, since these two parameters have little effect on the velocities. From the DS observations we obtained $v_{rot}$ sin i values of 39 km/s and 38 km/s for the primary (hotter and more massive star) and secondary of AP And, with estimated uncertainties of 3 km/s. However, the lower signal-to-noise ratios of these spectra, along with their shorter wavelength coverage and the considerable rotational line broadening, prevented us from determining the effective temperatures from this material. We therefore relied on the TRES spectra to select the templates. The TRES spectra yielded preliminary effective temperatures of 6580 ± 150 K and 6480 ± 150 K, and $v_{rot} \sin i$ values of 41 and 42 km/s, respectively, with estimated errors of 2 km/s. TRES and DS radial velocities were obtained using template parameters in our library nearest to these values (Teff = 6500 K and $v_{rot} \sin i$ = 40 km/s for both stars). For the DS spectra we applied corrections to the raw velocities following Torres et al. (1997), to account for possible systematic effects due to residual line blending and to lines shifting in and out of the narrow spectral window as a function of orbital phase. These corrections can be as large as 3 km/s for AP And, and they increase the masses by about 6%. Finally, the zero point of our velocity system for the DS instrument was monitored by taking nightly exposures at dusk and dawn, and small run-to-run corrections were applied as described by Latham (1992). For TRES we observed velocity standards every night. The final heliocentric radial velocities from both instruments with all corrections included are reported in Table 2 (and also in a machine-readable version with the electronic edition of this journal).

The DS and TRES spectra were also used to derive the light ratio between the stars as prescribed by Zucker & Mazeh (1994). We obtained similar values of $L_B/L_A = 0.95 \pm 0.06$ (DS) and $L_B/L_A = 0.93 \pm 0.02$ (TRES) at the mean wavelength of the observations (5187 Å).

Fekel et al. (2009) have provided a general description of the velocity measurement for our Fairborn Observatory echelle spectra. The solar-type star line list, consisting of mostly neutral lines, provided a better match to the spectra than our A-type star line list, which primarily consists of singly-ionized lines. Thus, from the former line list we measured about 165 lines of each component for radial velocity, fitting those lines with rotational broadening functions (Lacy & Fekel 2011) that allowed both the width and depth of the line fits to vary. Our unpublished measurements of several IAU solar-type velocity standards show that the Fairborn Observatory velocities have a zero-point offset of -0.6 km/s when compared to the results of Scarfe (2010). So, +0.6 km/s has been added to each velocity. The final measurements are listed in Table 2 (below, and also in a machine-readable version with the electronic edition of this journal), and have estimated uncertainties of about 2.0 and 1.7 km/s for star A and star B, based on the scatter from spectroscopic orbital fit described below.

From the Fairborn Observatory spectra, the average projected rotational velocities of the two stars are 40 and 39 km/s for the star A and star B, respectively, with estimated uncertainties of 2 km/s. The spectroscopic light ratio of the two components is $L_B/L_A = 0.90 \pm 0.02$ at an average wavelength of about 6000 Å.

Table 2: Heliocentric radial velocities for AP And.

| HJD-2400000 | $RV_A$(km/s) | $RV_B$(km/s) | $Err_A$(km/s) | $Err_B$(km/s) | Phase* |
|---|---|---|---|---|---|
| CfA/DS | | | | | |
| 53013.6038 | -46.82 | 46.22 | 1.74 | 2.42 | 0.4391 |
| 53038.5939 | -112.16 | 113.98 | 1.88 | 2.62 | 0.1829 |
| 53213.9698 | 107.62 | -113.52 | 1.44 | 2.00 | 0.6705 |
| 53218.8795 | 124.75 | -123.52 | 1.67 | 2.32 | 0.7636 |
| 53272.7785 | 123.60 | -127.59 | 1.45 | 2.02 | 0.7202 |
| 53280.6949 | 121.48 | -127.13 | 1.84 | 2.56 | 0.7076 |
| 53302.7597 | 77.00 | -78.15 | 1.86 | 2.58 | 0.6085 |
| 53308.6629 | -111.73 | 110.48 | 1.64 | 2.28 | 0.3275 |
| 53333.7138 | -80.45 | 76.34 | 1.57 | 2.18 | 0.1097 |
| 53335.6740 | -102.07 | 101.64 | 1.58 | 2.20 | 0.3446 |
| 53629.8372 | 107.86 | -111.39 | 1.65 | 2.29 | 0.6687 |
| 53656.8321 | 107.96 | -112.98 | 1.51 | 2.10 | 0.6755 |
| 53695.6998 | -104.03 | 102.04 | 1.75 | 2.44 | 0.1624 |
| 54457.6122 | -108.57 | 114.13 | 2.29 | 3.19 | 0.1703 |
| 54718.7922 | 119.57 | -123.88 | 1.56 | 2.17 | 0.7148 |
| 54742.7201 | 118.78 | -126.30 | 2.32 | 3.22 | 0.7895 |
| CfA/TRES | | | | | |
| 55134.6845 | 119.15 | -128.88 | 1.34 | 1.44 | 0.7287 |
| 55141.7537 | -114.16 | 114.64 | 1.20 | 1.28 | 0.1823 |
| 55168.7425 | -113.94 | 114.90 | 1.15 | 1.23 | 0.1854 |
| 55169.6299 | 122.23 | -125.61 | 1.28 | 1.37 | 0.7444 |
| 55195.6621 | -101.14 | 101.24 | 1.58 | 1.70 | 0.1448 |
| 55375.9744 | 118.31 | -127.15 | 2.35 | 2.52 | 0.7423 |
| 55382.9724 | -100.61 | 102.23 | 1.00 | 1.07 | 0.1511 |
| 55402.9170 | 118.97 | -122.80 | 0.89 | 0.95 | 0.7163 |
| 55515.7224 | 118.93 | -124.81 | 1.26 | 1.35 | 0.7841 |
| 55527.6464 | -117.00 | 118.59 | 1.01 | 1.09 | 0.2963 |
| 55584.5735 | -106.96 | 103.69 | 1.24 | 1.33 | 0.1606 |
| 55588.6043 | 115.54 | -119.41 | 1.05 | 1.12 | 0.7000 |
| 55811.8872 | -88.26 | 90.55 | 2.54 | 2.73 | 0.3692 |
| 55883.7718 | 101.33 | -105.70 | 0.64 | 0.69 | 0.6568 |
| 55884.7270 | -126.96 | 123.32 | 1.94 | 2.08 | 0.2586 |
| 55903.6981 | -120.72 | 121.16 | 1.14 | 1.22 | 0.2104 |
| 55961.6229 | 116.04 | -120.27 | 0.71 | 0.77 | 0.7033 |
| 55962.5761 | -115.71 | 119.43 | 1.07 | 1.15 | 0.3038 |
| 56090.9455 | -109.83 | 111.81 | 1.10 | 1.18 | 0.1771 |
| 56137.9719 | 113.71 | -120.05 | 1.38 | 1.49 | 0.8039 |
| 56172.8689 | 118.26 | -122.02 | 0.95 | 1.02 | 0.7892 |
| 56206.8282 | -113.37 | 114.64 | 0.80 | 0.86 | 0.1836 |
| 56237.7178 | 96.29 | -99.92 | 1.01 | 1.08 | 0.6442 |
| 56253.6398 | 108.53 | -112.54 | 0.72 | 0.78 | 0.6752 |
| 56257.5940 | -107.40 | 106.51 | 0.87 | 0.94 | 0.1663 |
| 56315.6212 | 120.36 | -125.53 | 0.78 | 0.84 | 0.7237 |
| FAIRBORN | | | | | |
| 55862.6476 | -97.7 | 102.8 | | | 0.3484 |
| 55873.6573 | -120.8 | 120.1 | | | 0.2846 |

| | | | |
|---|---|---|---|
| 55883.6426 | 54.9 | −60.1 | 0.5754 |
| 55933.6503 | −62.1 | 60.1 | 0.0804 |
| 56067.9586 | 113.9 | −119.0 | 0.6952 |
| 56070.9471 | 61.7 | −60.8 | 0.5780 |
| 56072.9490 | 105.1 | −105.1 | 0.8392 |
| 56074.9434 | −71.8 | 71.5 | 0.0957 |
| 56080.9274 | 90.8 | −95.9 | 0.8656 |
| 56086.9086 | 92.2 | −94.4 | 0.6338 |
| 56090.8970 | −100.6 | 98.9 | 0.1465 |
| 56093.8930 | −29.6 | 29.2 | 0.0340 |
| 56099.8850 | 117.4 | −115.6 | 0.8090 |
| 56101.8642 | −45.6 | 46.0 | 0.0559 |
| 56186.9350 | 99.6 | −103.9 | 0.6509 |
| 56209.6336 | 34.8 | −39.5 | 0.9511 |
| 56214.6844 | −93.4 | 93.5 | 0.1331 |
| 56238.6023 | −117.7 | 117.7 | 0.2015 |
| 56245.5984 | 80.5 | −81.4 | 0.6090 |
| 56264.5850 | 50.6 | −52.0 | 0.5707 |
| 56267.7174 | 36.4 | −34.8 | 0.5441 |
| 56273.6985 | −112.9 | 113.3 | 0.3122 |
| 56446.9189 | −46.7 | 41.7 | 0.4418 |
| 56457.8923 | −96.3 | 99.3 | 0.3551 |
| 56460.8852 | −123.8 | 123.6 | 0.2406 |
| 56461.8842 | 90.5 | −93.3 | 0.8700 |
| 56468.8587 | −124.1 | 123.1 | 0.2639 |
| 56550.7923 | 85.8 | −85.6 | 0.8825 |
| 56560.8882 | −124.6 | 122.4 | 0.2429 |
| 56564.7707 | 110.4 | −115.3 | 0.6889 |
| 56565.7818 | −108.3 | 111.5 | 0.3259 |
| 56566.7703 | 38.1 | −39.4 | 0.9487 |
| 56568.9329 | −112.9 | 114.8 | 0.3111 |
| 56569.8891 | 65.0 | −66.4 | 0.9135 |
| 56570.8891 | 35.5 | −36.4 | 0.5435 |
| 56572.6804 | 106.5 | −110.9 | 0.6721 |
| 56575.7422 | 73.9 | −75.6 | 0.6010 |
| 56576.7431 | −122.5 | 122.6 | 0.2316 |
| 56577.8684 | 45.1 | −49.4 | 0.9405 |
| 56578.8680 | 55.0 | −57.8 | 0.5703 |
| 56582.8591 | −64.3 | 65.7 | 0.0847 |

\* Based on the ephemeris in Sect. 2.

4. Spectroscopic Orbital Solution

Separate orbital fits were performed using the Levenberg-Marquardt method with each of the three data sets to check for systematic differences, holding the period and reference epoch fixed at the values determined in Sect. 1 and assuming the orbit to be circular. The elements are presented in Table 3, and are seen to be consistent with each other. The largest spread is in the semi-amplitude of the more massive star, $K_A$, which is marginally lower for TRES. The DS velocities show a slight zero-point offset between the star A and star B velocities (star A − star B = $\Delta RV$ = +1.56 ± 0.75 km/s), which is significant at the 2-sigma level. We accounted for this offset in the corresponding fit. The final solution was obtained by merging the three data sets, and using TRES as the reference group. We allowed for offsets between TRES and the other two data sets in addition to the star A – star B offset for the DS set, ΔRV. The uncertainties for the individual observations, which determine their relative weights, were rescaled by iterations to achieve a reduced chi-square near unity, separately for each group of velocities and for each component. This yields more realistic values for the uncertainties of the fitted parameters. The results of this combined fit are reported in the last column of Table 3. A graphical representation of the best fit together with the observations and residuals is shown in Figure 1.

The residuals from the Fairborn observations display some trends as a function of orbital phase that are also present in a solution that uses only those velocities. These trends are not apparent in the other data sets, which, however, are smaller in size and have a narrower phase coverage. It is possible the pattern is related to the line list used for the Fairborn reductions combined with the significant line broadening of both stars and partial blending of some of the measured lines with satellite lines from the other component in a phase-dependent way. We find, though, that removing the velocities within 50 km/s of the systemic velocity, which would be the ones most affected by blending, does not improve the situation. In any case these effects appear not to have a significant impact on the results, as the velocity semi-amplitudes are not very different from those derived with TRES and the DS.

Table 3: Spectroscopic orbital solutions for AP And*.

|  | CfA / DS | CfA / TRES | Fairborn | Combined fit |
|---|---|---|---|---|
| γ (km/s) | −0.29 ± 0.43 | −0.95 ± 0.15 | −0.64 ± 0.21 | −0.98 ± 0.15 |
| Star A−star B ΔRV (km/s) | +1.56 ± 0.74 | ... | ... | +1.83 ± 0.73 (DS) |
| ΔRV (TRES−DS) (km/s) | ... | ... | ... | −0.89 ± 0.45 |
| ΔRV (TRES−Fairborn) (km/s) | ... | ... | ... | −0.32 ± 0.26 |
| $K_A$ (km/s) | 123.92 ± 0.50 | 122.74 ± 0.22 | 123.60 ± 0.45 | 123.04 ± 0.19 |
| $K_B$ (km/s) | 125.93 ± 0.70 | 125.66 ± 0.23 | 125.36 ± 0.37 | 125.60 ± 0.19 |
| $q = M_B/M_A$ | 0.9840 ± 0.0067 | 0.9768 ± 0.0025 | 0.9860 ± 0.0047 | 0.9796 ± 0.0021 |
| a sin i (solar radii) | 7.839 ± 0.027 | 7.7938 ± 0.0099 | 7.811 ± 0.018 | 7.8012 ± 0.0083 |
| $M_A \sin^3 i$ (solar masses) | 1.293 ± 0.015 | 1.2752 ± 0.0052 | 1.2779 ± 0.0090 | 1.2770 ± 0.0044 |
| $M_B \sin^3 i$ (solar masses) | 1.272 ± 0.012 | 1.2456 ± 0.0049 | 1.2600 ± 0.0099 | 1.2510 ± 0.0042 |
| Span (days) | 1729.1 | 1180.9 | 720.2 | 2248.6 |
| N (star A/star B) | 16 / 16 | 26 / 26 | 41 / 41 | 83 / 83 |
| $rms_A$ (km/s) | 1.47 | 0.96 | 2.01 | 1.70 / 1.07 / 2.11 |
| $rms_B$ (km/s) | 2.28 | 1.06 | 1.70 | 2.37 / 1.14 / 1.76 |

*Note: The orbital period and epoch have been fixed at the photometric values in equation (1): HJD Min I = 2,454,717.65759(2) + 1.587291156(33) E. The projected semimajor axis and minimum masses rely on the solar radius and heliocentric gravitational constant adopted by Torres et al. (2010). Star A is the more massive, larger and more luminous star. Star B is the less massive, smaller, and less luminous star. ΔRV are radial velocity offsets due to differences in zero points.

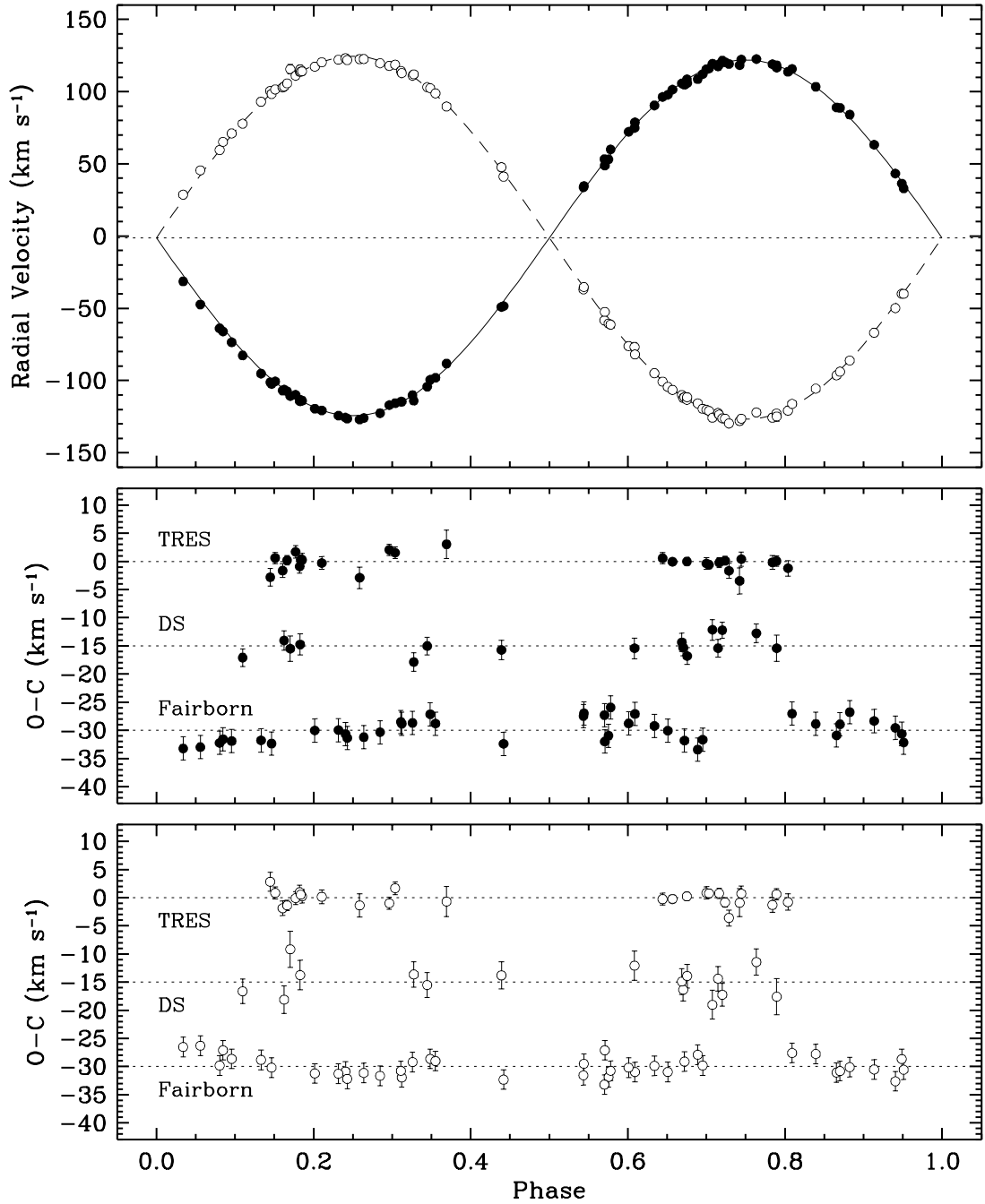

* Caption to Figure 1: (Top): Radial velocity observations for AP And along with our best-fit model from the combined solution. Filled symbols are for star A (the more massive, hotter star), and open symbols for the star B. The dotted line

represents the center-of-mass velocity. (Bottom panels): Velocity O-C residuals from the best fit (same symbols as above), with the DS and Fairborn values shifted vertically for clarity by -15 km/s and -30 km/s, respectively. In this diagram, phase 0.0 corresponds to a time of primary eclipse.

5. Photometric data and orbit

Sets of V-band differential photometry were obtained from images taken by two independent robotic telescopes, the URSA WebScope at the University of Arkansas campus and the NFO WebScope near Silver City, NM, USA. The URSA WebScope is constructed from a 10-inch diameter f/6.3 Meade LX200 Schimdt-Cassegrain telescope with an SBIG ST8 CCD camera binned 2x2 to produce 765x510 pixel images with 2.3 arcsec square pixels, inside a Technical Innovations Robo-Dome, all controlled by a computer inside a room at the top of Kimpel Hall on the Fayetteville campus. The NFO WebScope is constructed from a Group 128 24-inch diameter classical Cassegrain telescope with a Kodak KAF 4300E CCD camera producing 2102x2092 pixel images with 0.78 arcsec pixels. Both telescopes used Bessel V filters consisting of 2.0 mm of GG495 and 3.0 mm of BG 39. Exposures were 120 seconds long, and the cadence was typically 150 seconds per image. The star was observed with URSA from 2003 July 11 to 2012 July 13 (7892 images) and with the NFO from 2004 December 17 to 2013 December 2 (11205 images). The images were dark-corrected and flat-fielded from twilight flats, and in the case of the NFO, were corrected with a photometric flat (see Grauer et al. 2008). The images were measured with an application (Measure) written by one of the authors (CHSL). It automatically found the star pattern formed by the variable star and its comparison stars, corrected for differential airmass effects, and sky-subtracted the measurements of the stars to produce a differential magnitude between the eclipsing binary (var) and the magnitude corresponding to the sum of the fluxes of the two comparison stars, var-comps (comp 1 = TYC 3639-0767-1, Tycho $V_T$=11.69 (A5:), comp 2 = TYC 3639-1492-1, Tycho $V_T$=12.27 (A0:)). A few clearly errant observations were eliminated before analysis, but these observations are included in the machine-readable versions of the tables for completeness. The original measurements are listed in Table 4 (URSA) and Table 5 (NFO).

Table 4. V-band Differential Photometry (variable-comps) of AP And from the URSA WebScope*

| Orbital Phase | ΔV (mag) | HJD-2,400,000 |
|---|---|---|
| 0.96291 | 0.065 | 52,831.89683 |

| 0.96412 | 0.085 | 52,831.89874 |
| 0.96532 | 0.102 | 52,831.90065 |
| 0.96651 | 0.109 | 52,831.90254 |
| 0.96769 | 0.106 | 52,831.90441 |

(This table is available in its entirety in a machine-readable form in the online journal. A portion is shown here for guidance regarding its form and content.)
* Note: The orbital phase has been computed from the photometric values in equation (1): HJD Min I = 2,454,717.65759(2) + 1.587291156(33) E.

Table 5. V-band Differential Photometry (variable-comps) of AP And from the NFO WebScope*

| Orbital Phase | $\Delta V$ (mag) | HJD-2,400,000 |
|---|---|---|
| 0.49941 | 0.692 | 53,356.55449 |
| 0.50005 | 0.693 | 53,356.55551 |
| 0.50073 | 0.679 | 53,356.55658 |
| 0.50138 | 0.691 | 53,356.55761 |
| 0.50205 | 0.682 | 53,356.55867 |

(This table is available in its entirety in a machine-readable form in the online journal. A portion is shown here for guidance regarding its form and content.)
* Note: The orbital phase has been computed from the photometric values in equation (1): HJD Min I = 2,454,717.65759(2) + 1.587291156(33) E.

Because of imprecise centering from night to night and variations in responsivity across the field of view, small variations (about 0.01 mag) in the differential magnitude zero points are seen in the measurements (see Lacy, Torres, & Claret 2008 for a discussion of this effect). The URSA WebScope suffers very much less from this effect than does the NFO WebScope. We have measured these variations by doing a preliminary orbital fit to the URSA photometry, then correcting the data of each telescope for their nightly shifts. The number of these nightly corrections is shown in Table 6. The photometric model used here is the Nelson-Davis-Etzel model (Nelson & Davis 1972, Popper & Etzel 1981, Southworth, Maxted, & Smalley 2004) as implemented in the *jktebop* code of John Southworth. The *jktebop* code uses the Levenberg-Marquardt method to minimize the sum-of-squares of the residuals. A good description of details of the NDE model, upon which the *jktebop* code is based, is given by Etzel (1980), and is available from that author. The data corrected for nightly shifts are shown in Figures 2-4.

The meaning of parameters of the NDE model listed in Table 6 are $J_B$, the unitless central surface brightness of the cooler star (star B) relative to the central surface brightness of the hotter star (star A); $r_A+r_B$, the unitless sum-of-stellar-radii relative to the semi-major axis of the orbit (the radii correspond to those of spheres with volumes equivalent to the bi-axial ellipsoids used in the model); k, the unitless ratio of radii, $r_B/r_A$; $u_A$ and $u_B$, the unitless linear limb-darkening coefficients; i, the orbital inclination in degrees; q, the unitless ratio of masses $m_B/m_A$ (taken from the spectroscopic orbit and not adjusted); $L_A$ and $L_B$, the unitless observed fluxes of the components relative to the sum of the observed stellar fluxes at the orbital first quadrature phase (these are passband-specific fluxes; for our data, V band); $L_3$, the unitless third light flux relative to the sum of $L_A+L_B$; $\sigma$, the standard error of the residuals from the orbital fit, in magnitudes; N, the number of observations that were fitted; and the number of Corrections that were applied to compensate for the nightly shifts in the zero point of the magnitude scale. The auxilliary quantities $\beta_1$, the unitless gravity darkening exponents (which are normally taken from theory and not adjusted), we adopt from Claret (1998) as 0.28 for each star.

We have experimented with non-linear limb-darkening laws, both the quadratic and the logarithmic (Claret 2000), and find that they do not fit the data as well as the linear laws do. The reason for this is unknown, but could be that the non-linear laws are not as appropriate for these stars as the linear laws are. There were some small differences in the fitted values obtained using the non-linear laws. The fitted values of the radii $r_A$ and $r_B$ with the best fitting logarithmic law, which we found to be superior to the quadratic one, were smaller by 0.4% than the linear law values with the URSA data. With the NFO data the value of $r_A$ was 0.3% smaller with the logarithmic law, and the value of $r_B$ was the same value as with the linear law. Also, the fitted values of the inclination were smaller with the logarithmic law by about 0.2 degrees. This outcome is consistent with previous results such as those of Lacy, Torres, & Claret (2008). Southworth, Brundtt, & Buzasi (2007) present an example where non-linear limb-darkening was justified in the case of much more accurate observations. To be conservative, however, we have increased the uncertainty estimates of the adopted radii by 0.4% added in quadrature and the inclination uncertainty by 0.2 degrees added in quadrature.

The parameters of the fitted orbits that result in the curves in Figures 2-4 are given in Table 6. The increase in the apparent width of the observed points just before and after eclipses is simply due to the increased density of observations there, not to an increase in the standard error of the residuals. Tests were made to detect any orbital eccentricity or third light, but the values were insignificant, so were fixed at zero in the final fits. A test was done to see if the solution was sensitive to the value of the gravity darkening exponent $\beta_1$ by changing it from 0.28 to 0.32, but the difference was insignificant due to the fact that this system is well-detached, so the components are very close to spherical. Comparison of fitted parameters from the two independent data sets (URSA & NFO) is excellent, but to be conservative, half of the difference between the parameter values from the two telescopes was added in quadrature to the adopted parameter errors.

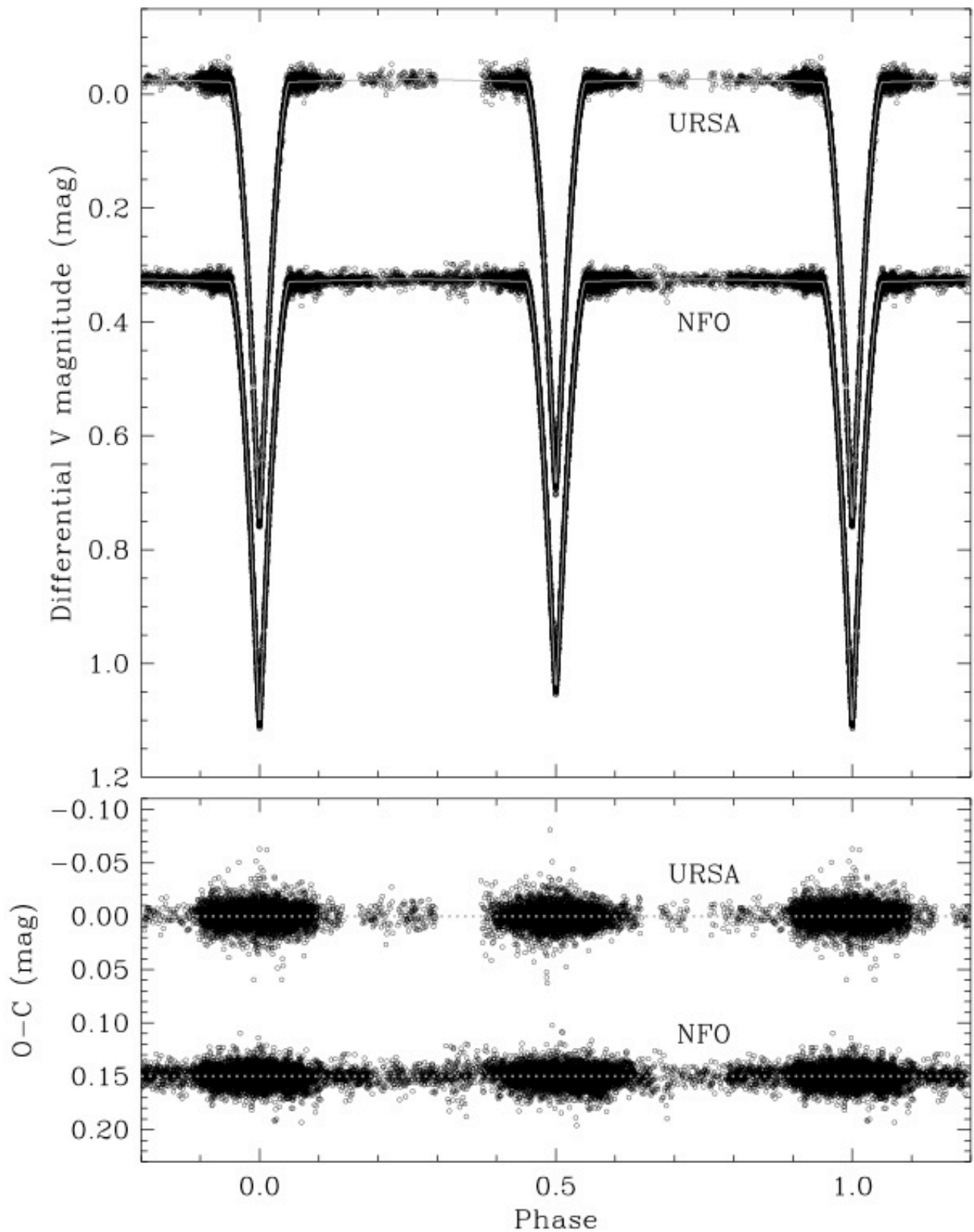

Figure 2. Light curves of AP And in the V-band from the URSA and NFO WebScopes. Residuals from the fitted model are shown in the bottom panel.

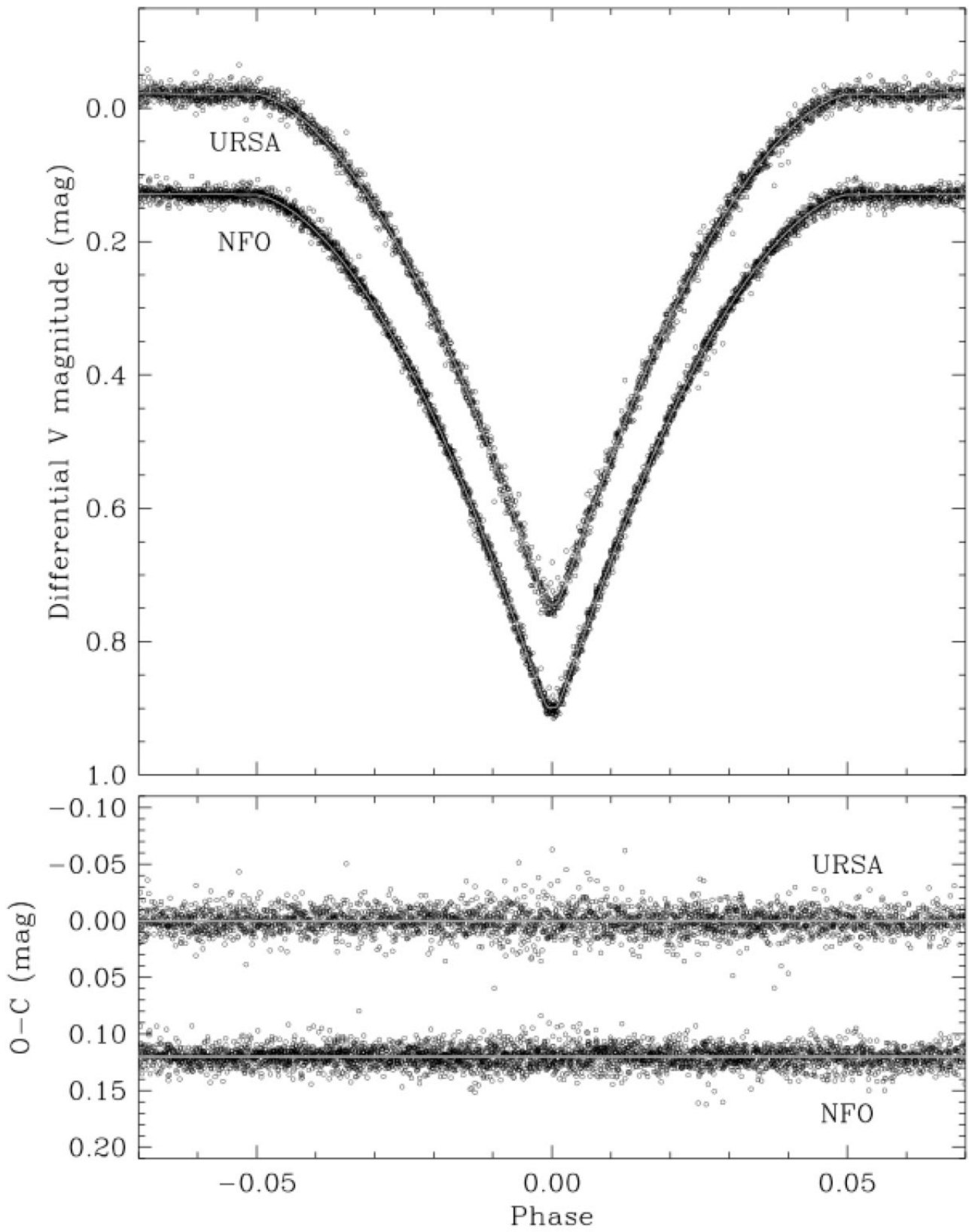

Figure 3. Primary eclipse of AP And.

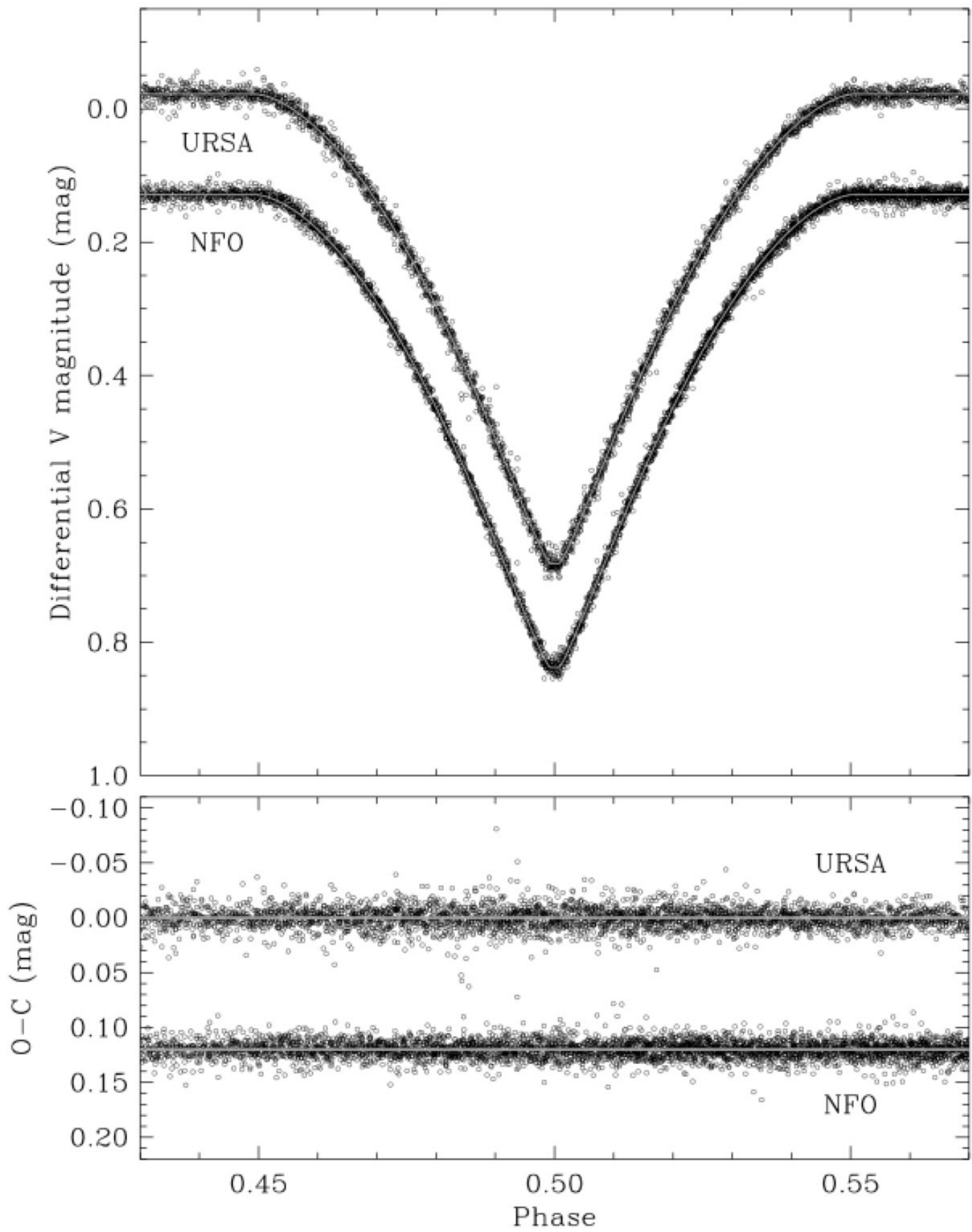

Figure 4. Secondary eclipse of AP And.

Table 6. Photometric light curve parameters of AP And*.

| Parameters | URSA | NFO | Adopted |
|---|---|---|---|
| $J_B$ | $0.9768 \pm 0.0067$ | $0.9691 \pm 0.0048$ | $0.9730 \pm 0.0077$ |
| $r_A+r_B$ | $0.31158 \pm 0.00021$ | $0.31122 \pm 0.00014$ | $0.31133 \pm 0.00029$ |
| $k$ | $0.9677 \pm 0.0016$ | $0.9708 \pm 0.0009$ | $0.9692 \pm 0.0022$ |
| $i$ (degrees) | $89.805 \pm 0.060$ | $89.915 \pm 0.069$ | $89.86 \pm 0.22$ |
| $u_A$ | $0.391 \pm 0.012$ | $0.448 \pm 0.008$ | $0.422 \pm 0.031$ |
| $u_B$ | $0.451 \pm 0.012$ | $0.481 \pm 0.009$ | $0.465 \pm 0.020$ |
| $q$ | 0.9796 fixed | 0.9796 fixed | 0.9796 fixed |
| $L_3$ | 0 fixed | 0 fixed | 0 fixed |
| $r_A$ | $0.15835 \pm 0.00016$ | $0.15791 \pm 0.00010$ | $0.15812 \pm 0.00078$ |
| $r_B$ | $0.15323 \pm 0.00016$ | $0.15331 \pm 0.00009$ | $0.15329 \pm 0.00066$ |
| $L_B/L_A$ | $0.893 \pm 0.015$ | $0.901 \pm 0.010$ | $0.897 \pm 0.016$ |
| $L_A$ | $0.5283 \pm 0.0011$ | $0.5261 \pm 0.0007$ | $0.5272 \pm 0.0016$ |
| $L_B$ | $0.4719 \pm 0.0070$ | $0.4740 \pm 0.0048$ | $0.4730 \pm 0.0073$ |
| $\sigma$ (mmag) | 9.93970 | 8.15293 | |
| N | 7892 | 11110 | |
| Corrections | 118 | 283 | |

* Note: The orbital period and epoch have been fixed at the photometric values in equation (1): HJD Min I = 2454717.65759(2) + 1.587291156(33) E.

6. Absolute dimensions

The combination of the spectroscopic and photometric elements leads to the accurate absolute properties for AP And listed in Table 7. Relative errors in the masses and radii are 0.3% and 0.5%, respectively, which are among the best for any known eclipsing binary. The temperature difference is determined much more precisely from the difference in eclipse depths than is implied by the formal uncertainties in the absolute temperatures, and is $70 \pm 25$ K. The more massive star is slightly hotter. We are not aware of any detailed study of the chemical composition of the components. Other derived properties for the stars are given also in Table 7, including the distance, 400 ± 30 pc, which relies on the visual absolute flux calibration of Popper (1980). A similar value is obtained when using bolometric corrections by Flower (1996). For the distance calculations we used an average apparent V-band magnitude for AP And of 11.15 ± 0.05 (Henden et al. 2012; Droege et al. 2006; and Hog et al. 2000 converted to the Johnson system), along with an extinction correction A(V) = 3.1 E(B-V). Estimates of the color excess E(B-V) were made using the reddening maps of Burstein & Heiles (1982), Schlegel et al. (1998), Drimmel et al. (2003), and Amores & Lepine (2005), and are 0.063, 0.057, 0.052, and 0.059 mag, respectively. We adopted the straight average along with a conservative error, 0.058 ± 0.030 mag.

There is excellent agreement in our $v_{rot} \sin i$ measurements for the components from three spectroscopic instruments. The weighted averages for each star, 40.2 ± 1.3 km/s and 40.0 ± 1.3 km/s, are consistent with the expected projected synchronous velocities listed in Table 7. The V-band light ratio from our photometric fits is also consistent with the estimates we obtained at similar wavelengths from our spectra, which supports the accuracy of our independent light-curve solutions, and in particular of the radii.

Table 7: Absolute dimensions of AP And

| Parameter | Star A | Star B |
|---|---|---|
| Mass (solar masses) | 1.2770 ± 0.0044 | 1.2510 ± 0.0042 |
| Radius (solar radii) | 1.2335 ± 0.0062 | 1.1954 ± 0.0053 |
| Log g (cgs) | 4.3623 ± 0.0044 | 4.3805 ± 0.0038 |
| Temperature (K) | 6565 ± 150 | 6495 ± 150 |
| Temperature difference (K) | 70 ± 25 | |
| Log L (solar units) | 0.403 ± 0.040 | 0.358 ± 0.039 |
| $F_V$* | 3.815 ± 0.011 | 3.810 ± 0.011 |
| $M_V$ (mag)* | 3.65 ± 0.11 | 3.78 ± 0.11 |
| $M_{bol}$ (mag)** | 3.723 ± 0.099 | 3.836 ± 0.099 |
| E(B-V) (mag) | 0.058 ± 0.030 | |
| m-M (mag)* | 8.02 ± 0.15 | |
| Distance (pc)* | 400 ± 30 | |
| Distance (pc)** | 390 ± 30 | |
| Measured $v_{rot} \sin i$ (km/s) | 40.2 ± 1.3 | 40.0 ± 1.3 |
| Synchronous $v_{rot} \sin i$ (km/s) | 39.3 ± 0.2 | 38.1 ± 0.2 |

* Relies on the visual absolute flux ($F_V$) calibration of Popper (1980), and is unitless.

** Relies on the absolute bolometric magnitude of the Sun from Torres (2010), and for the distance also on bolometric corrections of $BC_V = +0.01$ mag for both stars from Flower (1996), with a conservative error of 0.10 mag.

7. Comparison with stellar evolution models

The accurate absolute dimensions derived for AP And allow for a meaningful test of current stellar evolution models. Figure 5 shows our measurements against models from the Yonsei-Yale series of Yi et al. (2001), for a best-fit metallicity of $Z = 0.0150$, corresponding to [Fe/H] = -0.09 in these models. The stars are seen to lie squarely on the zero-age main sequence, with a best-fit age of only 500 Myr. The agreement with the models is excellent. In particular, the measured temperature difference, which relies on the difference in the eclipse depths, is well reproduced by theory.

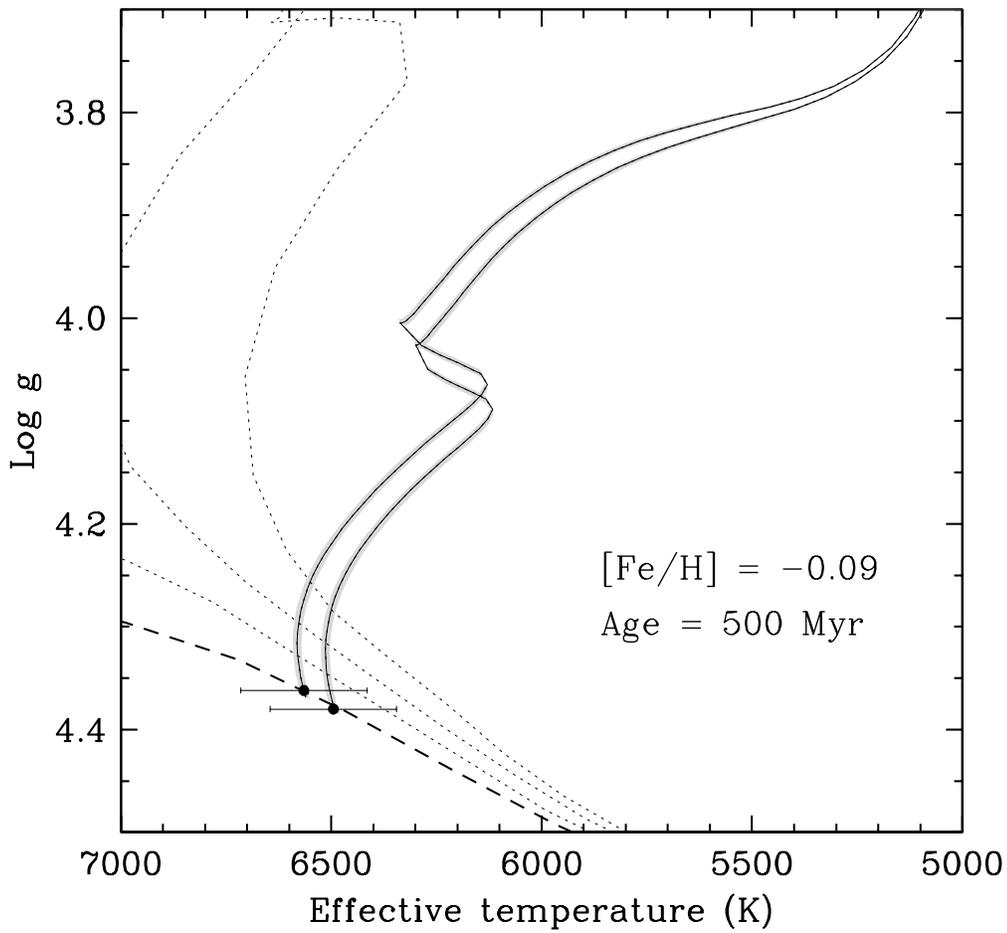

* Caption to Figure 5: Observations for AP And compared against stellar evolution models by Yi et al. (2001) for a metallicity of [Fe/H] = -0.09 and [α/Fe] = 0.0. Here α/Fe is the enrichment factor for nuclear α-capture elements such as O, Ne, Mg, etc. relative to iron (Kim et al. 2002). Evolutionary tracks are shown with solid lines for the exact masses we measure. The shaded areas represent the uncertainty in the location of the tracks that comes from the

observational errors in the masses. The best fit 500 Myr isochrone is indicated with a dashed line, and other isochrones (1, 1.5, and 2 Gyr) are shown with dotted lines.

Among the eclipsing binaries with accurately measured properties in this mass range (Torres et al. 2010) AP And is the least evolved. The system of V505 Per (Tomasella et al. 2008) is a near clone, with primary and secondary masses that are less than 0.4% different from those of AP And and temperatures nominally only 35-55 K cooler, but larger radii such that the age is a factor of three older. Unevolved systems such as AP And constitute excellent tests of models regarding the metallicity, though unfortunately a spectroscopic abundance analysis is lacking for AP And.

ACKNOWLEDGMENTS

The authors wish to thank Bill Neely who operates and maintains the NFO WebScope for the Consortium, and who handles preliminary processing of the images and their distribution.  Thanks also to University of Arkansas undergraduate student Craig Heinrich for initial analysis of the URSA photometry and preliminary radial velocities. We thank P. Berlind, Z. Berta, M. L. Calkins, G. A. Esquerdo, G. Furesz, D. W. Latham, R. P. Stefanik, and S. Tang for help with the spectroscopic observations of AP And on Mount Hopkins, as well as R. J. Davis and J. Mink for maintaining the echelle databases at the CfA. GT acknowledges partial support through NSF grant AST-1007992. The research at Tennessee State University was made possible by NSF support through grant 1039522 of the Major Research Instrumentation Program.  In addition, astronomy at Tennessee State University is supported by the state of Tennessee through its Centers of Excellence programs.  We wish to thank an anonymous referee of this article for suggestions that improved the clarity of our paper and expanded some of our discussion.